\documentclass[a4paper]{jpconf}
\usepackage{graphicx}

\begin{document}

\title{Investigation of light and heavy tetraquark candidates using lattice QCD}

\author{Marc Wagner$^1$, Abdou Abdel-Rehim$^{2}$, Constantia Alexandrou$^{2,3}$, Mattia Dalla Brida$^4$, Mario Gravina$^{3}$, Giannis Koutsou$^{2}$, \\ Luigi Scorzato$^5$, Carsten Urbach$^6$}

\address{$^1$ Goethe-Universit\"at Frankfurt am Main, Institut f\"ur Theoretische Physik, \\ $\phantom{xxx}$Max-von-Laue-Stra{\ss}e 1, D-60438 Frankfurt am Main, Germany}
\address{$^2$ Computation-based Science and Technology Research Center, Cyprus Institute, 20 Kavafi \\ $\phantom{xxx}$ Street, Nicosia 2121, Cyprus}
\address{$^3$ Department of Physics, University of Cyprus, P.O.\ Box 20537, 1678 Nicosia, Cyprus}
\address{$^4$ School of Mathematics, Trinity College Dublin, Dublin 2, Ireland}
\address{$^5$ INFN -- gruppo collegato di Trento, via sommarive 14, 38123 -- Trento, Italy}
\address{$^6$ Helmholtz-Institut f{\"u}r Strahlen- und Kernphysik (Theorie) and Bethe Center for \\ $\phantom{xxx}$ Theoretical Physics, Universit{\"a}t Bonn, D-53115 Bonn, Germany}

\ead{mwagner@th.physik.uni-frankfurt.de}

\begin{abstract}
We review the status of an ongoing long-term lattice investigation of the spectrum and structure of tetraquark candidates. We focus on the light scalar meson $a_0(980)$. First steps regarding the study of a possibly existing $c c \bar{c} \bar{c}$ tetraquark are also outlined.
\end{abstract}


\section{Introduction}

The nonet of light scalar mesons formed by $\sigma \equiv f_0(500)$, $\kappa \equiv K_0^\ast(800)$, $a_0(980)$ and $f_0(980)$ is poorly understood \cite{Jaffe:2004ph}. Compared to expectations based on a standard quark antiquark picture  all nine states are rather light and their ordering is inverted. This can, however, naturally be explained assuming a tetraquark structure, which is also supported by certain decay channels, e.g.\ $a_0(980) \rightarrow K + \bar{K}$.

Here we report about the status of an ongoing long-term project with the aim to study such tetraquark candidates using lattice QCD. We mainly focus on the $a_0(980)$. In section~\ref{SEC001} we summarize recently published results obtained with Wilson twisted mass quarks, where diagrams with closed fermion loops (also called ``singly disconnected diagrams'') have been neglected \cite{Daldrop:2012sr,Alexandrou:2012rm,Wagner:2012ay,Wagner:2013nta,Wagner:2013jda}. In section~\ref{SEC455} we discuss latest technical advances obtained with clover improved Wilson quarks, in particular the inclusion of diagrams with closed fermion loops. Finally, in section~\ref{SEC769} we discuss first steps of a study of a possibly existing $c c \bar{c} \bar{c}$ tetraquark, which has recently been predicted using a coupled system of covariant Bethe-Salpeter equations \cite{Heupel:2012ua}.


\section{\label{SEC001}Wilson twisted mass quarks, diagrams with closed fermion loops neglected}


\subsection{\label{SEC388}Lattice setup}

We use gauge link configurations with 2+1+1 dynamical quark flavors generated by the ETM Collaboration \cite{Baron:2010bv,Baron:2010th}. We consider several ensembles with lattice spacing $a \approx 0.086 \, \textrm{fm}$. The ensembles differ in the volume $(L/a)^3 \times (T/a) = 20^3 \times 48 , \ldots , 32^3 \times 64$ and the unphysically heavy $u/d$ quark mass corresponding to $m_\pi \approx 280 \, \textrm{MeV} \ldots 460 \, \textrm{MeV}$.

For the computations presented in this section we have ignored diagrams with closed fermion loops, which are technically rather challenging (cf.\ section~\ref{SEC500}). Physical consequences are discussed in detail in \cite{Alexandrou:2012rm}.


\subsection{Four-quark creation operators}

$a_0(980)$ has quantum numbers $I(J^P) = 1(0^+)$. As usual in lattice QCD we extract the low lying spectrum in that sector by studying the asymptotic exponential behavior of correlation functions $C_{j k}(t) = \langle (\mathcal{O}_j(t))^\dagger \mathcal{O}_k(0) \rangle$ \cite{Weber:2013eba}. $\mathcal{O}_j$ and $\mathcal{O}_k$ denote suitable creation operators, i.e.\ operators generating the $a_0(980)$ quantum numbers, when applied to the vacuum state.

Assuming that the experimentally measured $a_0(980)$ with mass $980 \pm 20 \, \textrm{MeV}$ is a rather strongly bound four quark state, suitable creation operators to excite such a state are
\begin{eqnarray}
\label{EQN001} & & \hspace{-0.7cm} \mathcal{O}_{a_0(980)}^{K \bar{K} \textrm{\scriptsize{} molecule}} \ \ = \ \ \sum_\mathbf{x} \Big(\bar{s}(\mathbf{x}) \gamma_5 u(\mathbf{x})\Big) \Big(\bar{d}(\mathbf{x}) \gamma_5 s(\mathbf{x})\Big) \\
\label{EQN002} & & \hspace{-0.7cm} \mathcal{O}_{a_0(980)}^{\textrm{\scriptsize diquark}} \ \ = \ \ \sum_\mathbf{x} \Big(\epsilon^{a b c} \bar{s}^b(\mathbf{x}) C \gamma_5 \bar{d}^{c,T}(\mathbf{x})\Big) \Big(\epsilon^{a d e} u^{d,T}(\mathbf{x}) C \gamma_5 s^e(\mathbf{x})\Big) .
\end{eqnarray}
The first operator has the spin/color structure of a $K \bar{K}$ molecule. The second resembles a bound diquark antidiquark pair, where spin coupling via $C \gamma_5$ corresponds to the lightest diquarks/antidiquarks (cf.\ e.g.\ \cite{Jaffe:2004ph,Alexandrou:2006cq,Wagner:2011fs}). Further low lying states with $a_0(980)$ quantum numbers, which need to be considered are the two-particle states $K + \bar{K}$ and $\eta_s + \pi$. Suitable creation operators to resolve these states are
\begin{eqnarray}
\label{EQN003} & & \hspace{-0.7cm} \mathcal{O}_{a_0(980)}^{K + \bar{K} \textrm{\scriptsize{} two-particle}} \ \ = \ \ \bigg(\sum_\mathbf{x} \bar{s}(\mathbf{x}) \gamma_5 u(\mathbf{x})\bigg) \bigg(\sum_\mathbf{y} \bar{d}(\mathbf{y}) \gamma_5 s(\mathbf{y})\bigg) \\
\label{EQN004} & & \hspace{-0.7cm} \mathcal{O}_{a_0(980)}^{\eta_s + \pi \textrm{\scriptsize{} two-particle}} \ \ = \ \ \bigg(\sum_\mathbf{x} \bar{s}(\mathbf{x}) \gamma_5 s(\mathbf{x})\bigg) \bigg(\sum_\mathbf{y} \bar{d}(\mathbf{y}) \gamma_5 u(\mathbf{y})\bigg) .
\end{eqnarray}


\subsection{\label{SEC377}Numerical results}

\begin{figure}[htb]
\begin{center}
\includegraphics[angle=-90,width=7.0cm]{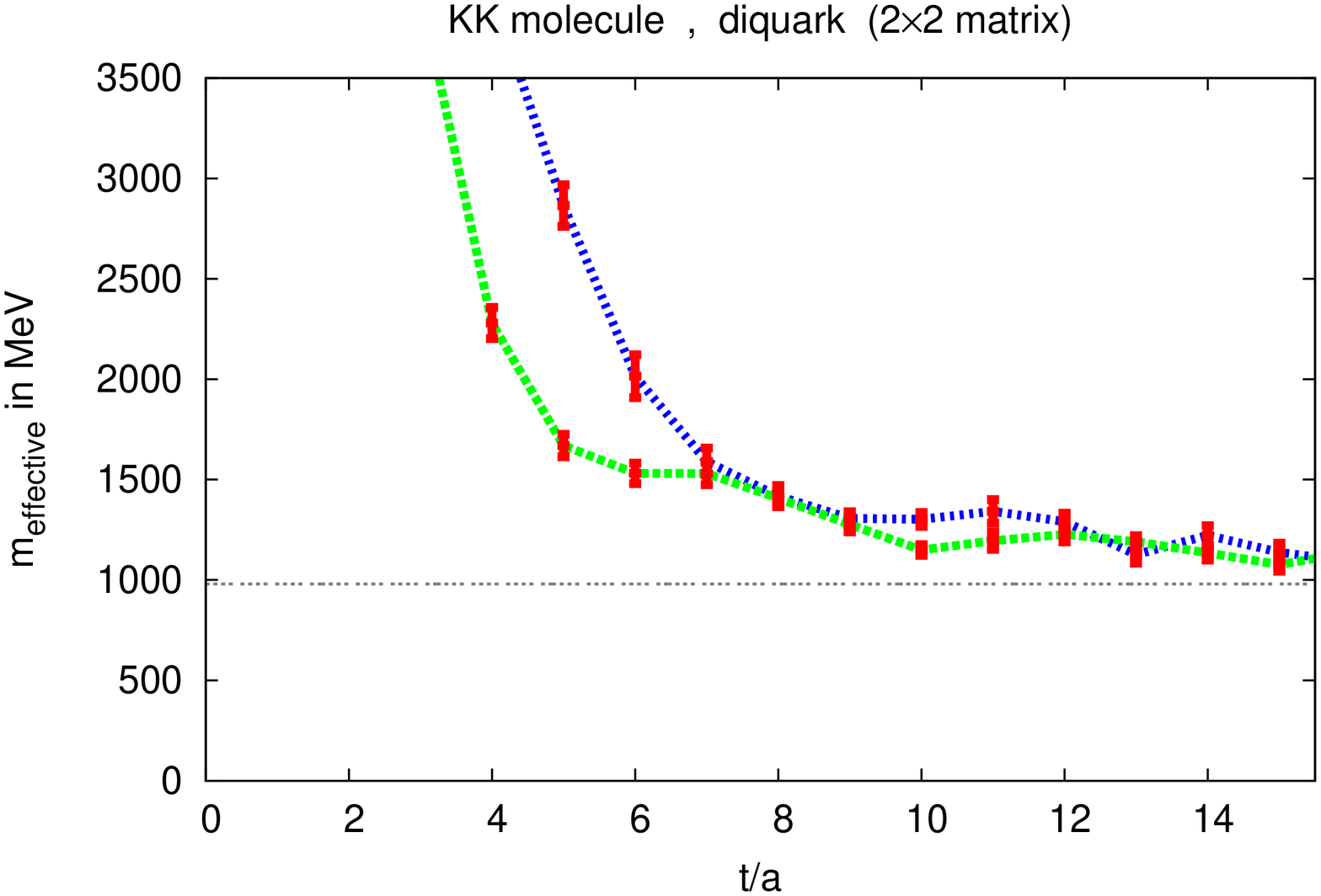}
\includegraphics[angle=-90,width=7.0cm]{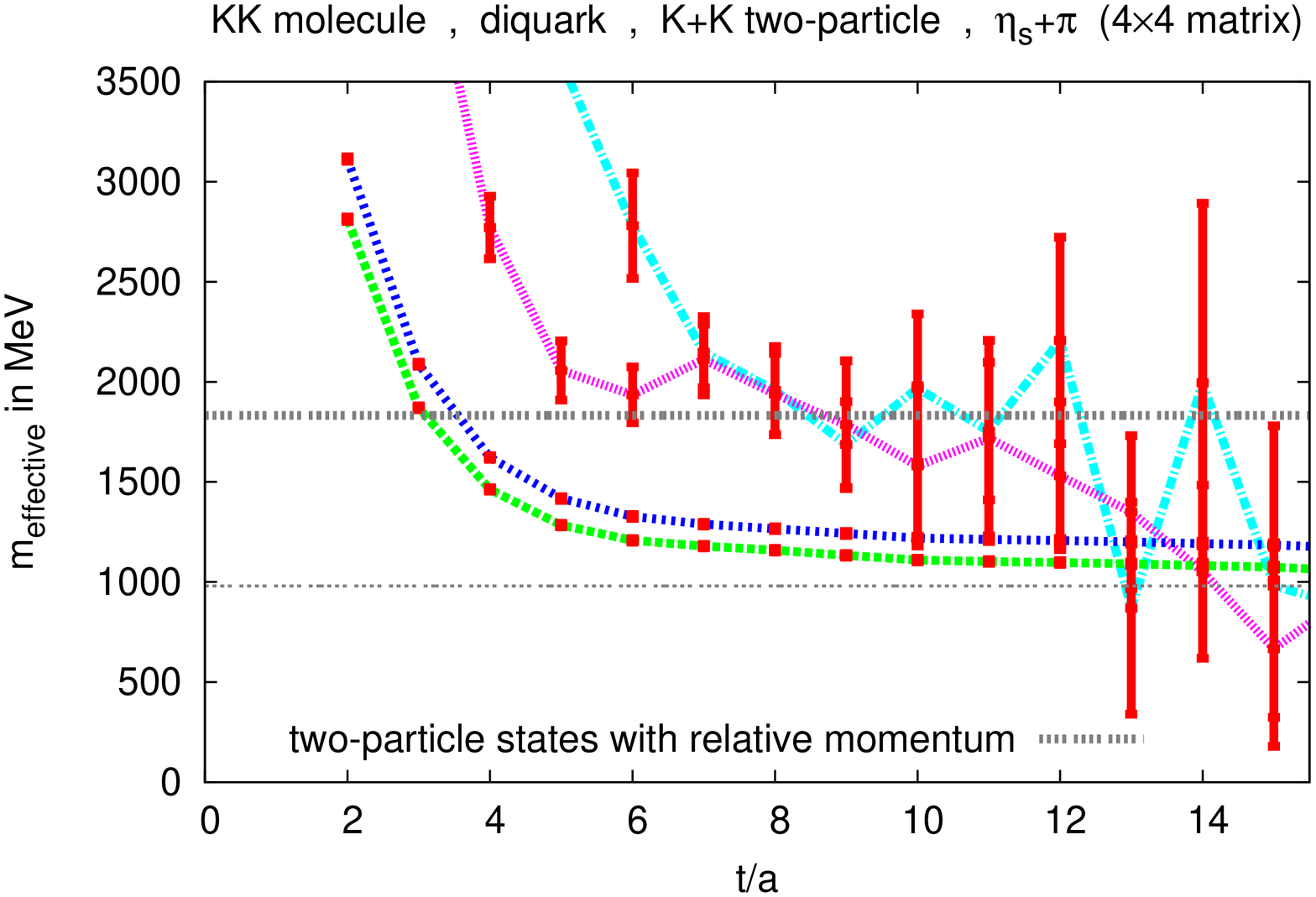}
\caption{\label{F001}Effective masses as functions of the temporal separation.
\textbf{(left)}~$2 \times 2$ correlation matrix (operators: $K \bar{K}$ molecule, diquark).
\textbf{(right)}~$4 \times 4$ correlation matrix (operators: $K \bar{K}$ molecule, diquark, $K + \bar{K}$ two-particle, $\eta_s + \pi$ two-particle).
}
\end{center}
\end{figure}

Figure~\ref{F001}(left) shows effective mass plots for a $2 \times 2$ correlation matrix with a $K \bar{K}$ molecule operator (\ref{EQN001}) and a diquark-antidiquark operator (\ref{EQN002}) (lattice volume $(L/a)^3 \times (T/a) = 20^3 \times 48$, pion mass $m_\pi \approx 341 \, \textrm{MeV}$). The corresponding two plateaus are around $1100 \, \textrm{MeV}$ and, therefore, consistent both with the expectation for possibly existing $a_0(980)$ tetraquark states and with two-particle $K + \bar{K}$ and $\eta_s + \pi$ states, where both particles are at rest ($2 \times m(K) \approx 1198 \, \textrm{MeV}$; $m(\eta_s) + m(\pi) \approx 1115 \, \textrm{MeV}$ in our lattice setup).

Increasing this correlation matrix to $4 \times 4$ by adding $K + \bar{K}$ two-particle and $\eta_s + \pi$ two-particle operators (eqs.\ (\ref{EQN003}) and (\ref{EQN004})) yields the effective mass results shown in Figure~\ref{F001}(right). Two additional states are observed, whose plateaus are around $1500 \, \textrm{MeV} \ldots 2000 \, \textrm{MeV}$.

From these plots one can conclude that the two low-lying states are the expected two-particle $K + \bar{K}$ and $\eta_s + \pi$ states, while an additional stable $a_0(980)$ tetraquark state does not exist. The second and third excitation can also be interpreted as two-particle states, which have non-vanishing relative momentum. A detailed discussion of these plots and corresponding arguments can be found in \cite{Alexandrou:2012rm}.

Qualitatively identical results are found, when using other ensembles with light quark masses and spacetime volumes as mentioned in section~\ref{SEC388}.


\section{\label{SEC455}Clover improved Wilson quarks, diagrams with closed fermion loops included}


\subsection{Lattice setup}

Recently we started to perform similar computations using gauge link configurations generated by the PACS-CS Collaboration \cite{Aoki:2008sm} with 2+1 flavors of clover improved Wilson sea quarks, a lattice spacing $a \approx 0.09 \, \textrm{fm}$, a volume $(L/a)^3 \times T/a = 32^3 \times 64$ and a $u/d$ quark mass corresponding to $m_\pi \approx 300 \, \textrm{MeV}$. A significant advantage compared to Wilson twisted mass quarks is that parity and isospin are exact symmetries. For example there is no pion and kaon mass splitting and $P = +$ and $P = -$ states are separated by quantum numbers (these problems in the context of Wilson twisted mass quarks and the $a_0(980)$ are discussed in detail in \cite{Alexandrou:2012rm}).


\subsection{\label{SEC500}Diagrams with closed fermion loops}

A major improvement compared to our previous results presented in section~\ref{SEC377} is that this time diagrams with closed fermion loops are included, i.e.\ where strange quark propagators start and end at the same timeslice. Ignoring such diagrams introduces a systematic error, which is hard to quantify. With these new computations this systematic error has been eliminated. Another important consequence of the inclusion of closed fermion loop is that quark-antiquark and four-quark trial states can have non-vanishing overlap. This allows to study a larger correlation matrix containing not only the four-quark operators (\ref{EQN001}) to (\ref{EQN004}), but also a quark-antiquark operator
\begin{eqnarray}
\label{EQN645} \mathcal{O}_{a_0(980)}^{q \bar{q}} \ \ = \ \ \sum_\mathbf{x} \Big(\bar{d}(\mathbf{x}) u(\mathbf{x})\Big) .
\end{eqnarray}
Such a correlation matrix will enable us to make stronger statements about the structure of states from the $a_0(980)$ sector. The techniques we use to compute these diagrams with closed fermion loops are described in \cite{Wagner:2013jda}.


\subsection{First numerical results}

\begin{figure}[htb]
\begin{center}
\includegraphics[width=7.0cm]{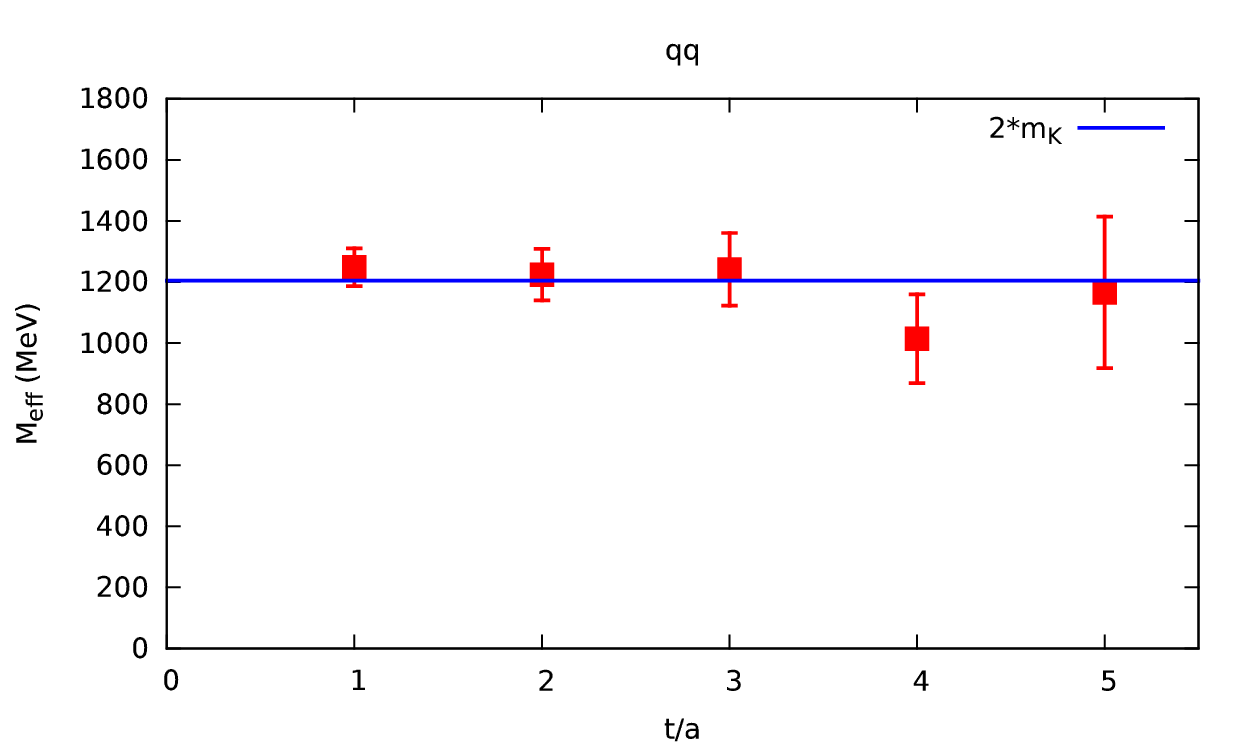}
\includegraphics[width=7.0cm]{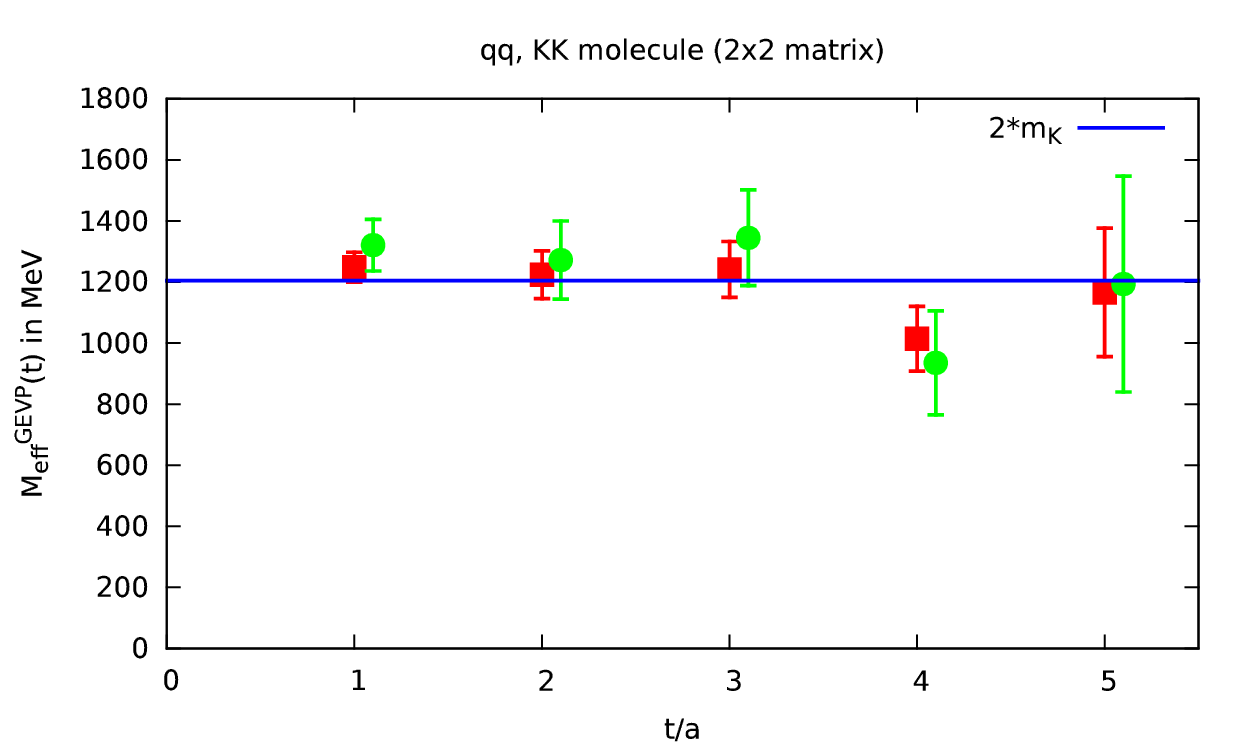}
\caption{\label{F004}Effective masses as functions of the temporal separation.
\textbf{(left)}~$q \bar{q}$ operator.
\textbf{(right)}~$2 \times 2$ correlation matrix (operators: $q \bar{q}$, $K \bar{K}$ molecule).
}
\end{center}
\end{figure}

At the moment only the computation of a $2 \times 2$ correlation matrix containing the $q \bar{q}$ operator (eq.\ (\ref{EQN645})) and the four-quark $K \bar{K}$ molecule operator (eq.\ (\ref{EQN001})) has been finished. The effective masses corresponding to the diagonal element of the $q \bar{q}$ operator is plotted in Figure~\ref{F004}, left. The plot on the right shows the two effective masses obtained by solving a generalized eigenvalue problem using the full $2 \times 2$ correlation matrix. All effective masses are around $1000 \, \textrm{MeV}$ and, therefore, consistent with $2 \times m(K)$, with $m(\eta) + m(\pi)$ and also with $m(a_0(980))$.

For statements regarding the structure of the observed states we need to include also the four-quark operators (\ref{EQN002}) to (\ref{EQN004}), which is currently in progress. This will allow us to address similar questions as in section~\ref{SEC377}, i.e.\ to answer, whether there is in addition to the two-particle $K + \bar{K}$ and $\eta + \pi$ states also a third (bound) state near the mass of $a_0(980)$. We also intend to study the volume dependence of the two-particle spectrum using ``L\"uscher's method'' (cf.\ e.g.\ \cite{Luscher:1986pf,Luscher:1990ux,Luscher:1991cf}) or improved techniques taking e.g.\ coupled channel scattering into account \cite{Bernard:2010fp,Doring:2012eu,Doring2013}. Such computations are very challenging using lattice QCD, but first results have recently been published, e.g.\ for the $\kappa$ and positive parity $D$ mesons \cite{Lang:2012sv,Mohler:2012na}.


\section{\label{SEC769}Lattice investigation of a possibly existing $\bar{c} c \bar{c} c$ tetraquark}

Recently a $\bar{c} c \bar{c} c$ tetraquark has been predicted using a coupled system of covariant Bethe-Salpeter equations \cite{Heupel:2012ua}. It is expected to be a mesonic molecule-like state formed by two $\eta_c$ mesons with a mass of $m(\bar{c} c \bar{c} c) = (5.3 \pm 0.5) \, \textrm{GeV}$. Comparing this mass to $2 \times m(\eta_c) = 6.0 \, \textrm{GeV}$ indicates that the binding energy could be quite large, $\Delta E = m(\bar{c} c \bar{c} c) - 2 \times m(\eta_c) \approx -(0.7 \pm 0.5) \, \textrm{GeV}$. The uncertainty quoted in \cite{Heupel:2012ua} is, however, of the same order of magnitude. Investing such a $\bar{c} c \bar{c} c$ tetraquark using lattice methods could provide an independent confirmation of its existence.

We use the same setup as discussed for the $a_0(980)$ meson in section~\ref{SEC455}. In a first attempt we consider a single creation operator of molecule type, which models a bound $\eta_c \eta_c$ state,
\begin{eqnarray}
\label{EQN650} \mathcal{O}_{\bar{c} c \bar{c} c}^{\eta_c \eta_c \textrm{\scriptsize{} molecule}} \ \ = \ \ \sum_\mathbf{x} \bigg(\bar{c}(\mathbf{x}) \gamma_5 c(\mathbf{x})\bigg) \bigg(\bar{c}(\mathbf{x}) \gamma_5 c(\mathbf{x})\bigg) .
\end{eqnarray}
Diagrams with closed fermion loops are ignored.

\begin{figure}[htb]
\begin{center}
\includegraphics[width=7.0cm]{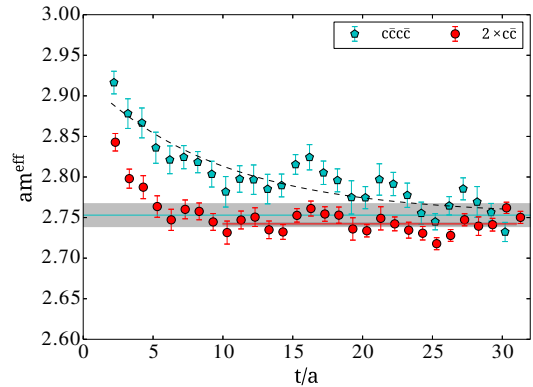}
\caption{\label{F005}Effective mass as functions of the temporal separation for the $\eta_c \eta_c$ molecule operator (blue points) and the $\eta_c$ meson multiplied by a factor $2$ (red points).}
\end{center}
\end{figure}

In Figure~\ref{F005} we compare the corresponding effective mass (blue points) with two times the $\eta_c$ effective mass (red points). We do not observe any indication of a four quark $\bar{c} c \bar{c} c$ state, which is lighter than $2 \times m(\eta_c)$. On the other hand the effective mass corresponding to the creation operator (\ref{EQN650}) still seems to slightly decrease at large temporal separations and does not exhibit a clear plateau. This signals a trial state $\mathcal{O}_{\bar{c} c \bar{c} c}^{\eta_c \eta_c \textrm{\scriptsize{} molecule}} | \Omega \rangle$, which has a poor ground state overlap. The ground state could either be two unbound $\eta_c$ (then the plateau is expected at $2 \times m(\eta_c)$) or still a $\bar{c} c \bar{c} c$ bound state, however, with a structure quite different from (\ref{EQN650}).

Currently we explore the latter alternative. While the creation operator (\ref{EQN650}) places the two $\eta_c$ mesons essentially on top of each other, they could be rather far separated, e.g.\ by a distance up to $r \approx 1 \, \textrm{fm}$ (for a recent study of this distance for heavy-heavy-light-light tetraquarks cf.\ \cite{Bicudo:2012qt}). To this end we employ corresponding more general molecule type $\bar{c} c \bar{c} c$ creation operators. Computations are in progress.


\section*{Acknowledgments}

We acknowledge helpful discussion with Christian Fischer. M.W.\ acknowledges support by the Emmy Noether Programme of the DFG (German Research Foundation), grant WA 3000/1-1. M.G.\ acknowledges support by the Marie-Curie European training network ITN STRONGnet grant PITN-GA-2009-238353. M.D.B.\ is funded by the Irish Research Council and is grateful for the hospitality at the University of Cyprus and Cyprus Institute, where part of this work was carried out. L.S.\ acknowledges support by STRONGnet. This work was supported in part by the Helmholtz International Center for FAIR within the framework of the LOEWE program launched by the State of Hesse and by the DFG and the NSFC through funds provided to the Sino-German CRC 110 ``Symmetries and the Emergence of Structure in QCD''.



\end{document}